\documentclass[%
 aip,
 jcp,%
 amsmath,amssymb,
 preprint,%
]{revtex4-1}

\usepackage{graphicx}
\usepackage{dcolumn}
\usepackage{amsmath,amssymb,bm}
\usepackage{color}

\begin{document}


\title[Mechanism of wetting]{Mechanism of the
Cassie-Wenzel transition via the atomistic and continuum string methods}

\author{Alberto Giacomello}
\email{alberto.giacomello@uniroma1.it}
\altaffiliation[Also at ]{Max-Planck-Institut f\"ur Intelligente Systeme, 70569 Stuttgart, Germany}
\affiliation{Dipartimento di Ingegneria Meccanica e Aerospaziale, Universit\`a di Roma ``La Sapienza'', 00184 Rome, Italy}

\author{Simone Meloni}
\email{simone.meloni@epfl.ch}
\affiliation{Institute of Chemical Sciences and Engineering, \'Ecole
Polytechnique F\'ed\'erale de Lausanne, CH-1015 Lausanne, Switzerland}

\author{Marcus M\"uller}
\affiliation{Institut f\"ur Theoretische Physik,
Georg-August-Universit\"at G\"ottingen, 37077 G\"ottingen, Germany}

\author{Carlo Massimo Casciola}
\affiliation{Dipartimento di Ingegneria Meccanica e
Aerospaziale, Universit\`a di Roma ``La Sapienza'', 00184 Rome, Italy}

\date{\today}

\begin{abstract}
The string method is a general and flexible strategy to compute the most probable
transition path for an activated process (rare event). We apply here the
atomistic string method in the density field to the Cassie-Wenzel
transition, a central problem in the field of superhydrophobicity.  We
discuss in detail the mechanism of wetting of a submerged hydrophobic
cavity of nanometer size and its dependence on the geometry of the
cavity. Furthermore, we discuss the algorithmic analogies between the
string method and CREaM [Giacomello \emph{et al.}, Phys. Rev. Lett.
\textbf{109}, 226102 (2012)], a method inspired by the string that
allows for a faster and simpler computation of the mechanism and of the
free-energy profiles of the wetting process. This approach is general
and can be employed in mesoscale and macroscopic calculations.
\end{abstract}

\pacs{PACS}
\keywords{Cassie-Wenzel transition, superhydrophobicity, rare events, molecular dynamics, mechanism, string method}

\maketitle
\section{Introduction}

Wetting of chemically and topographically heterogeneous surfaces gives
rise to a rich phenomenology and a corresponding wealth of theoretical
challenges.\cite{Quere2008,rauscher2008} A remarkable example is that a careful
combination of surface roughness and chemistry yields highly
liquid-repellent and self-cleaning surfaces under given environmental
conditions: this class of properties is often referred to as
\emph{superhydrophobicity}.  Superhydrophobicity is related to the
trapping of gaseous pockets (air and/or vapor) inside surface
roughness.\cite{lafuma2003} We will loosely refer to this scenario as
the Cassie state. The superhydrophobic Cassie state also favors the
emergence of liquid slippage under flow conditions.\cite{tretyakov2013,gentili2014}
However, superhydrophobicity breaks down as soon as the
surface roughness becomes wet in the so-called Wenzel state.
As a consequence of the very different properties of the Cassie and
Wenzel states, there is a growing interest in designing surfaces that
are capable of stabilizing the superhydrophobic Cassie state. In order
for such engineering effort to be effective, a thorough knowledge of the
Cassie-Wenzel transition is required. With this objective, we analyze
here the mechanism of the Cassie-Wenzel transition with the string
method applied to molecular dynamics simulations.

The phase transition between the Cassie and the Wenzel states is, in most
cases, characterized by large free-energy
barriers.\cite{dupuis2005,koishi2009,savoy2012b,Giacomello2012,checco2014} The
superhydrophobic Cassie state on the same surface can be stable,
metastable \footnote{A metastable state is a local minimum of the free
energy  in which the system can be trapped, even for very long time,
because of the free-energy barriers separating it from the stable state
(absolute minimum).}, or unstable depending on the environmental conditions.
Therefore it is not correct to speak about \emph{superhydrophobic
surfaces}, but rather about \emph{superhydrophobic
states}.\cite{lafuma2003} The problem
of designing surfaces with superhydrophobic properties is therefore one
of maximizing the range of temperatures, pressures, and characteristics
of the liquids/vapor phases in contact for which the Cassie state is stable. 
In the conditions where the Cassie
state is not thermodynamically stable, it is nonetheless possible to
exploit metastabilities to obtain long-living superhydrophobicity: this
is the case, \emph{e.g.}, of omniphobic surfaces that present a
metastable Cassie state even with ``wetting'' liquids thanks to a
special reentrant geometry.\cite{tuteja2007} Indeed, free-energy
barriers must be much larger than the thermal energy $k_B T$ in order
for the metastable Cassie state to survive for times that are
significant for experiments and applications.  Knowledge of
how the Cassie-Wenzel transition starts and evolves --the transition
path or wetting mechanism-- may yield new insights for the design of engineered
surfaces.

According to the transition state theory, the rate of the Cassie-Wenzel
transition depends exponentially on the free-energy barrier between these two
states. Thus, designing surfaces with the desired Cassie-Wenzel free-energy
barrier is an effective tool for controlling the rate of the process. For
instance it has recently been suggested\cite{poetes2010} that the Cassie state is
generally metastable underwater. This statement, which is based on experiments
on a small number of surfaces, could be made more precise if the way in which
free-energy barriers depend on the geometry of the surface roughness was known.
The crucial point for applications is not whether the desired
state is metastable or stable but whether it will last longer than the
experiment/application. 
The first step in engineering surfaces is, therefore, the
characterization of the wetting mechanism and of its dependence on the
topography and chemistry of the surface, as well as on the thermodynamic
conditions. 

\begin{figure}
\centering
\includegraphics[width=9cm]{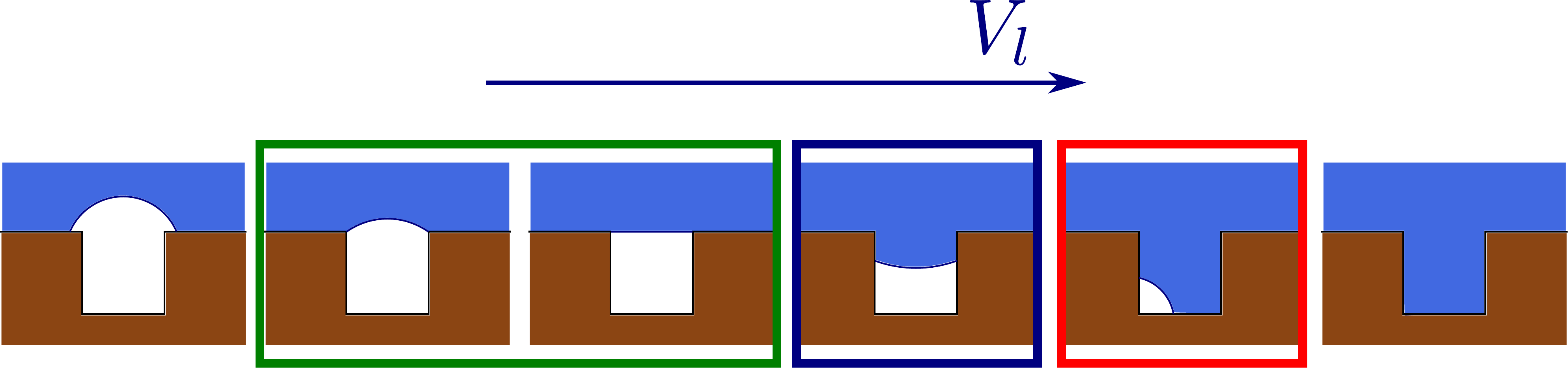}\\
\vspace{0.5cm}
\includegraphics[width=9cm]{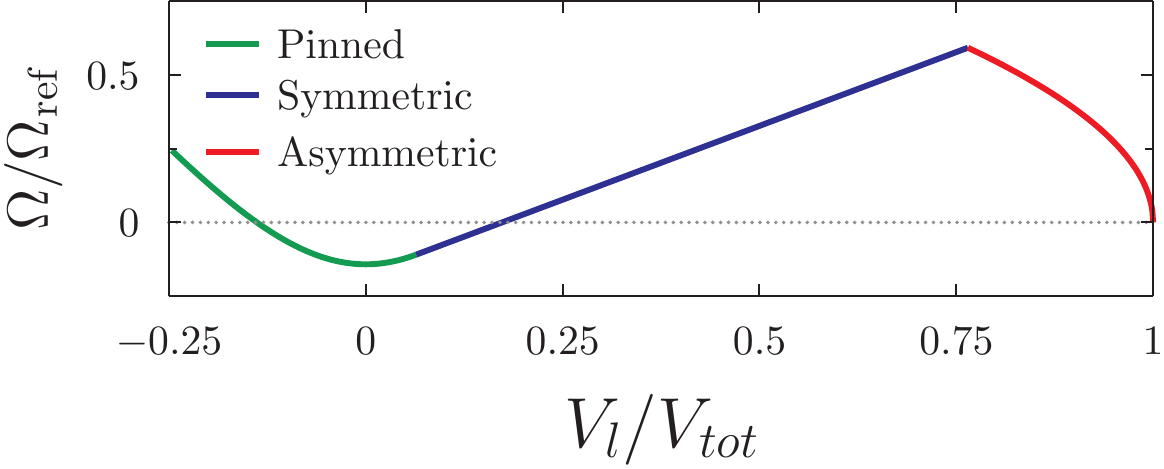}
	\caption{Mechanism of the Cassie-Wenzel transition on rectangular
		grooves as computed with the continuum rare events
	method\cite{Giacomello2012} (\emph{top}): the path is parameterized in
	terms of the volume of liquid filling the cavity, $V_l$. Free-energy profile of the
Cassie-Wenzel transition as computed from the path above at coexistence (\emph{bottom}). \label{fig:intro}}
\end{figure}

Previous works typically considered the mechanism of the Cassie-Wenzel transition on drops, see \emph{e.g.} Ref.~\onlinecite{kusumaatmaja2008,papadopoulos2013}, but there is a growing interest in submerged surfaces where the external pressure plays a key role in the stability of the superhydrophobic state. We focus here on the latter case by studying a model system that is simple enough to allow comparison of different approaches and yet shows a surprisingly rich phenomenology (see Fig.~\ref{fig:intro}).

In previous works
\cite{giacomello2012langmuir,Giacomello2012,giacomello2013} we
characterized the free-energy barriers of the Cassie-Wenzel transition
occurring in isolated hydrophobic roughness elements under different
conditions of pressure and temperature. The systems considered spanned from
few nanometers (explored \emph{via} molecular dynamics
simulations\cite{giacomello2012langmuir}) to macroscopic dimensions,
for which the Continuum Rare Events Methods, or
CREaM, was developed.\cite{Giacomello2012,giacomello2013} 
Over this broad range of systems, at coexistence --when the Cassie and
Wenzel state have the same free-energy--, free-energy barriers are
much larger than $k_BT$ accounting for strong metastabilities.  

For all previous approaches, the wetting path was characterized following the
changes in the (coarse-grained) density field of the fluid, $\rho(x)$.
This, in turn, was considered a parametric function of the filling
of the surface corrugation (or liquid volume inside it), $\rho(x;V_l)$. The
resulting path of the transition is the
sequence $\{\rho(x;V_{l,i})\}_{i=1,N}$ of density fields minimizing
the free-energy at a given $V_{l,i}$. Under suitable conditions,
explained in Sec.~\ref{sec:CREaM}, this represent a realistic description of the wetting path. 
However, when this description was applied in
combination with a macroscopic, sharp-interface macroscopic
model,\cite{Giacomello2012} we obtained a discontinuous wetting path
(see Fig.~\ref{fig:intro}). The discontinuity corresponds to an
instantaneous switching from a symmetric liquid/vapor meniscus to an
asymmetric bubble in one of the corners of the corrugation
(morphological transition). This discontinuity occurs at the
``transition state'' and results in a non-differentiability of the free
energy profile at this point.
This sharp point may have two distinct origins,
\begin{itemize}
\item an algorithmic one, related to the parameterization of the wetting
	path with the volume of liquid in the groove used in CREaM
\item a modeling one, that is, it could arise as a genuine feature of
	the sharp-interface model, which was used in combination with
	CREaM.\cite{Giacomello2012,giacomello2013}
\end{itemize}

The goal of this paper is therefore to address the question about the
sharp point of the free-energy profile and, more generally, to lay on
solid statistical grounds the discussion about the wetting mechanism on rough
surfaces. In order to achieve this goal, we compute the \emph{minimum
free-energy path} (MFEP) using the string method in collective
variables.\cite{maragliano2006} We employ atomistic simulations with the
aim of making minimal assumptions on the liquid/vapor interface and the
interactions with the walls. The collective variable that characterizes
the microscopic configuration implemented is
the coarse-grained density field.  
We compare atomistic with continuum, sharp-interface model paths and free-energy
profiles. The continuum path is obtained with the string and CREaM methods.
Qualitatively, the atomistic and continuum wetting are consistent.
Surprisingly, atomistic string free-energy profile shows a better agreement with
continuum CREaM. This is due to an ``error cancellation'', with CREaM
compensating intrinsic limitations in the sharp-interface model with respect to
the atomistic case.

The second goal of this work is to
elucidate the effect of the shape of the surface corrugation and its
size on the mechanism of the Cassie-Wenzel transition and on the related
free-energy barrier. Anticipating our results, the concept of
transition path itself may break down if the corrugations are sufficiently
small.

The paper is organized as follows: in Sections~\ref{sec:MD},
\ref{sec:Interface}, and \ref{sec:CREaM} the methods employed here, the
\emph{atomistic string}, the \emph{interface string}, and the
\emph{continuum rare events method} (CREaM) are introduced and compared. This
first part contains the main methodological findings. In
Section~\ref{sec:results} the atomistic and continuum results are presented and
discussed, concentrating on the physics of the Cassie-Wenzel transition. The
last section summarized all conclusions.

\section{Molecular Dynamics Simulations}
\label{sec:MD}

The mechanism of the Cassie-Wenzel transition was investigated with the
string method in collective variables applied to molecular dynamics
simulations\cite{maragliano2006}. Molecular dynamics simulations were
performed with the LAMMPS engine\cite{LAMMPS} equipped for the string
calculations with the PLUMED\cite{PLUMED} plugin as explained below.
The isothermal/isobaric ensemble (NPT) was used for all simulations by
using the algorithm of Martyna \emph{et
al.}\cite{martyna1992nose,martyna1994}.  The standard Lennard Jones (LJ)
potential was used for the fluid-fluid interactions; fluid-solid
interactions were also of LJ type, with the attractive term that was
tuned through the factor $c$:
\begin{equation}
	\Phi_{LJ} (r_{ij}) = \epsilon\left[\left(\frac{\sigma}{r_{ij}}\right)^{12} -
	c \left(\frac{\sigma}{r_{ij}}\right)^{6} \right] \text{ ,}
\end{equation}
where $r_{ij}$ is the distance between the atoms $i$ and $j$, while
$\epsilon$ and $\sigma$ set the scales of energy and length,
respectively.
In order to obtain a hydrophobic solid, we set $c=0.6$, which
corresponds to a contact angle of $\theta_Y\simeq110^\circ$. Periodic boundary conditions
are applied in the three directions.  The lower wall featured a
rectangular groove (or trench) extending through the $y$ direction.
Two kinds of grooves were considered,
the first having a width of $11\sigma$ and square aspect ratio and
the second having a width of $22\sigma$ and a rectangular aspect ratio,
see Fig.~\ref{fig:CV}.

\subsection{Coarse grained density field}
\begin{figure}
	\includegraphics[height=4cm]{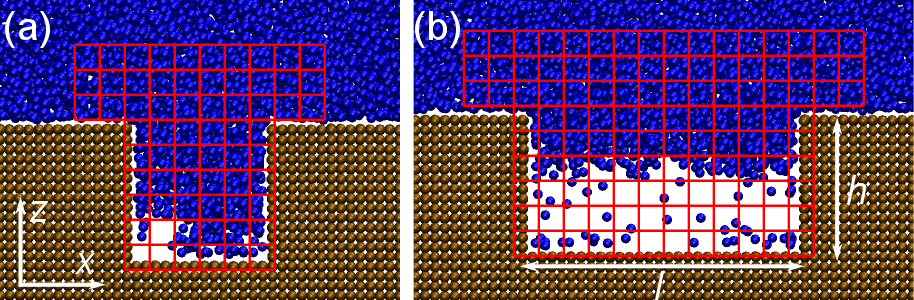}
	\caption{Cells used in the definition of the coarse-grained density
		collective variable for the square (a) and the rectangular groove (b). The axes
		and relevant dimensions used in the text are also defined. \label{fig:CV}}
\end{figure}

The collective variable used to describe the intrusion of
liquid inside of the groove was the coarse-grained density field. This
quantity was computed from the atomic positions as the number of atoms
inside the cells sketched in Fig.~\ref{fig:CV}. We used a mollified
version of the characteristic function of the cells based on the Fermi
functions in order to prevent impulsive forces on atoms crossing the
cell boundaries (see Sec.~\ref{sec:algorithm}). The cells occupied the whole $y$ dimension of the groove
thus being effectively two-dimensional: for the square groove a total of
$N=66$ cells were used, while for the rectangular groove $N=120$, as sketched
in Fig.~\ref{fig:CV}.

The Landau free-energy of the system as a function of the realization, $\bm z$, of
the (vector) collective variable $\bm \theta(\bm r)$ is defined as:
\begin{equation}
	F(\bm z) = - k_B T \ln P(\bm z) =
	- k_B T \ln \left(\int \mathrm d\bm r\;\;  \mathrm m(\bm r) \prod_{i=1}^N \delta(\theta_i(\bm r) -z_i)
	\; \right) \text{ ,}
\label{eq:landau}
\end{equation}
where $k_B$ is the Boltzmann constant, $T$ is the system temperature, $P(\bm z)$ is the probability to find the system at state $\bm z$, and $\bm r$ is the $3 N_p$ dimensional vector of particles positions, with $N_p$ the number of particles in the system. 
The probability $P(\bm z)$ is expressed in the second equality of
Eq.~\eqref{eq:landau} as the integral over the $3N_p$-dimensional configurational space of the
probability density $\mathrm{m}(\bm r)$ of the relevant ensemble (the Boltzmann factor)
times Dirac deltas centered at value $z_i$ of the $N$
components of the collective variable. The collective variable is
assumed to depend only on the $3N_p$ configurational degrees of freedom.

\subsection{Implementation of the string method}
\label{sec:algorithm}
For the general derivation of the string method in collective variables
we refer the reader to the original work of Maragliano \emph{et
al.}\cite{maragliano2006} Briefly, this method allows one to identify
the minimum free-energy path (MFEP), that is, the path of maximum
likelihood. The MFEP is the continuous 
curve in the space of collective variables  --in this case the
coarse-grained density field-- satisfying the equation  
\begin{equation}
	\frac{d z_i(\alpha)}{d \alpha} \quad \bigg\lvert \bigg\rvert \quad
	\sum_{j=1}^N M_{ij}(\bm z(\alpha)) \frac{\partial F(\bm z(\alpha))}{\partial z_j} \text{ ,}
	\label{eq:MFEP}
\end{equation}
where $\alpha$ is a parameterization of the MFEP, $\parallel$ means
``parallel to'', the indices $i$ and $j$ run over the $N$ collective variables
(which in vector notation are indicated as $\bm z$), $F(\bm z)$ is the
free-energy defined in Eq.~\eqref{eq:landau}, and  $M_{ij}(\bm z)$ is
a metric matrix due to projection of the phase space onto the collective variable space,
and defined as\cite{maragliano2006}
\begin{eqnarray}
	M_{ij}(\bm z) &=& 
	                     \langle {\boldsymbol \nabla}_{\bm r}\theta_i \cdot {\boldsymbol \nabla}_{\bm r}\theta_j \rangle_{\bm z} \nonumber \\
	&\equiv&  
	\frac{\int\; 	\mathrm d \,\bm r\;\;  
	{\boldsymbol \nabla}_{\bm r}\theta_i \cdot {\boldsymbol \nabla}_{\bm r}\theta_j
	\;\; \mathrm e^{-\beta U(\bm r)} \prod_{k=1}^N \delta(\theta_k(\bm r) -z_k)}
	{\int\; 	\mathrm d \,\bm r\;\;  
	\mathrm e^{-\beta U(\bm r)} \prod_{k=1}^N \delta(\theta_k(\bm r) -z_k)}
	  \text{ ,}
	\label{eq:metricmatrix}
\end{eqnarray}
where $\beta^{-1}=k_B T$ and $U(\bm r)$ is the potential energy of the
system.
Loosely speaking, when the metric matrix is unitary, Eq.~\eqref{eq:MFEP}
prescribes that the MFEP joins two minima of the free-energy landscape
passing through the bottom of the valleys and the saddle point connecting them (the
transition state).

The string method is an algorithm that allows one to identify the MFEP.
The string itself is a discretization of the path connecting two
metastable states, that is, two minima in the free-energy landscape. The
string is parameterized according to its relative arc-length,
$\alpha =
\int_{{\bm z}_a}^{{\bm z}_\alpha} 	|\mathrm d \bm z| /  \int_{{\bm
z}_a}^{{\bm z}_b} 	|\mathrm d \bm z|$, with $a$ and $b$ beginning and end
of the string.  The discrete points along the string are called
\emph{images} and are labeled with their position on the string
$\alpha_l$, where $l$ is the index of the images.  We use here the version of the string method by E \emph{et
al.}\cite{e2007}, which consists of three steps:
\begin{enumerate}
	\item Calculation of the free-energy gradient and of the metric
	matrix, see RHS of Eq.~\eqref{eq:MFEP}, at the current position of the
	images;
	\item Evolution of one timestep of the images according to the (time-discretized) pseudo-dynamics
\begin{equation}
	\frac{\partial z_i(\alpha_l,t)}{\partial t} =
	-\sum_{j=1}^N M_{ij}(\bm z(\alpha_l,t)) \frac{\partial F(\bm
	z(\alpha_l,t))}{\partial z_j} \text{ ;}
	\label{eq:string}
\end{equation}
	\item Parameterization of the string to enforce equal arc-length
		parameterization among contiguous images.
\end{enumerate}
For sufficiently large $t$, the algorithm guarantees that $z(\alpha_l,t)$ converges to the
MFEP of Eq.~\eqref{eq:MFEP}.~\cite{maragliano2006,e2007}

Steps 2 and 3 of the string algorithm above can be written also as\cite{maragliano2006}
\begin{equation}
\label{eq:stringOrtho}
\frac{d {\bm z}(\alpha_l, t)}{d t} = - {\hat M}(\bm z(\alpha_l, t)) \nabla
F(\bm z(\alpha_l, t)) \left(\bm 1 - \bm \tau(\bm z(\alpha_l, t)) \otimes \bm \tau (\bm z(\alpha_l, t))
	\right)\, ,
\end{equation}
where the term $\bm 1 - \bm \tau(\bm z(\alpha_l, t)) \otimes \bm \tau (\bm
z(\alpha_l, t))$ projects out the component of $- {\hat M}(\bm z(\alpha_l, t))
\nabla F(\bm z(\alpha_l, t))$ along the string. 
his projector, indeed, implements  the constraint $\alpha = const.$ in the dynamics of $z(\alpha_l, t)$ (see Eq.~2 of Ref.~\cite{e2007}).
Summarizing, the string method is a first order dynamics with the constraint of constant value of the parameter $\alpha_l$ following the generalized force $- {\hat M}(\bm z(\alpha_l, t)) \nabla
F(\bm z(\alpha_l, t))$. In other words, the string method is a  constrained (local) minimization of the system at a set of states at
fixed $\alpha_l$ in a space equipped with the metric $\hat M$.

In order to compute the free-energy gradient and the metric matrix
required to evolve the string \emph{via} Eq.~\eqref{eq:string} (point
$1.$ of the algorithm) we used
\emph{restrained molecular dynamics} (RMD, see Ref.~\onlinecite{maragliano2006}). In practice,
a restraining potential was added to the Hamiltonian of the system in the
form $\kappa/2(\theta_i(\bm r) -z_i)^2$, where $\theta_i(\bm r)$ is the current
value of the $i$-th component of the collective variable vector and
$z_i$ is its target value.  For sufficiently large $\kappa$, the
dynamics driven by the restrained Hamiltonian samples the conditional
ensemble at $\bm \theta(\bm r) =\bm z$, and allows one to compute the
relevant quantities at a given position of the string
\cite{maragliano2006}. 

\subsection{Transition state ensemble}
\label{sec:TS}
Once the string method has converged, it is possible to explore local
approximations to the isocommittor surfaces, which are the hyperplanes
orthogonal (\emph{via} the metric $\hat M$) to the string. In
particular, we considered here the hyperplane intersecting the string at
the transition state; this plane bears the relevant information about
the transition region, that is, the region where the probability to fall
into the products' side is the same as in the reactants'
\cite{maragliano2006}. The transition ensemble is a collection of
microscopic states with the constraint that the system lies on this plane. 

The \emph{transition state ensemble} just discussed was computed \emph{via} restrained
molecular dynamics implementing the prescription
\begin{equation}
	\sum_{i,j=1}^N \tau_i (\alpha_{TS}) M_{ij}\left (\bm z(\alpha_{TS})\right) (z_j - z_j(\alpha_{TS})) = 0
	\text{ ,}
\end{equation}
where $\bm \tau (\alpha_{TS})$ is the tangent to the string at
the transition state (TS). The configurations extracted from the transition
state ensemble are discussed in Sec.~\ref{sec:results}.

\section{The interface string}
\label{sec:Interface}
Rather than describing the configuration of the system by the coarse-grained
density field, $\rho(x,z)$, one can characterize the configuration of the
liquid inside the cavity by the position, $h_{i}$, of the liquid-vapor
interface.  Conceptually, one can obtain the corresponding Landau free-energy,
$\Omega[h_{i}]$, by Eq.~(\ref{eq:landau}) using the collective variable, $h=\bm
\theta(\bm r)$. This strategy requires a definition of the interface position,
$h$, in terms of the microscopic degrees of freedom, $\{{\bm r}\}$, which can
be achieved via the coarse-grained density field. 

In the following, however, we do not compute the Landau free-energy from the
particle simulations but, instead, we use the mean-field, sharp-interface
approximation to estimate $\Omega$. This grand potential takes the form
\cite{Giacomello2012}
\begin{equation}
\Omega = -p_{l}V_{l} - p_{v}V_{v} + \gamma_{lv}A_{lv}  + \gamma_{sv}A_{sv} + \gamma_{sl}A_{sl}
\label{eq:grand}
\end{equation}
where $V_{tot}$, $V_{l}$, and $V_{v}$ denote the total volume and the volumes
occupied by the liquid and the vapor, respectively. $\gamma_{lv}$,
$\gamma_{sl}$, and $\gamma_{sv}$ characterize the liquid-vapor interface
tension, and the surface tension of the confining solid with the liquid and the
vapor. $A_{lv}$, $A_{sl}$, and $A_{sv}$ are the contact areas of the three
coexisting phases.

Additionally, like in capillary-wave Hamiltonians or solid-on-solid
models, we
assume that the interface position, $z=h_{i}(x)$, is a single-valued function,
i.e., configurations with overhangs and bubbles are ignored. Using a completely
filled cavity as reference state,
$\Omega_{0}=-p_{l}V_{tot}+\gamma_{sl}l_{y}(2h+l)$ with $h$ being the height of
the cavity and $l$ its widths, we obtain for the excess grand potential $\Delta
\Omega =\Omega - \Omega_{0}$ 
\begin{eqnarray}
\Delta \Omega                                        &=& (p_{l}-p_{v})V_{v} + \gamma_{lv}A_{lv} + (\gamma_{sv}-\gamma_{sl})A_{sv} 
\label{eq:omega_contrib}\\
\frac{\Delta \Omega}{V_{tot}\Delta p_{\max}} &=& \Delta \tilde p \frac{V_{v}}{V_{tot}} - \frac{lA_{lv}}{2V_{tot}\cos \theta_{Y}} -  \frac{l A_{sv}}{2V_{tot}} 
\label{eq:omega_rescaled}
\end{eqnarray}
with the abbreviations $\Delta p_{\max} = - 2 \gamma_{lv} \cos \theta_{Y}/l$,
$\Delta \tilde p = \frac{p_{l}-p_{v}}{\Delta p_{\max}}$, and Young's equation
$\gamma_{lv}\cos \theta_{Y}=\gamma_{sv}-\gamma_{sl}$.

Next, we express the vapor volume, $V_{v}$, the interface area, $A_{lv}$, and
the solid-vapor contact area, $A_{sv}$, through the interface position,
$h_{i}(x)$
\begin{eqnarray}
\frac{V_{v}}{V_{tot}} &=& \frac{1}{lh} \int_{0}^{l} {\rm d}x\;h_{i}(x) \\
\frac{lA_{lv}}{V_{tot}} &=& \frac{1}{h} \int_{0}^{l} {\rm d}x\; \sqrt{1+\left(\frac{{\rm d}h_{i}}{{\rm d}x}\right)^{2}} \\
\frac{lA_{sv}}{V_{tot}} &=& \frac{h_{i}(0)+h_{i}(l)+l}{h}
\end{eqnarray}
The expressions will be appropriate, if the liquid-vapor interface is above the
bottom of the cavity, i.e., $h>0$. If the liquid-vapor interface touches the
bottom of the cavity, however, the liquid will be in contact with the solid,
$\gamma_{sl}$, rather than two separate liquid-vapor and vapor-solid
interfaces. We account for this effect by the correction term
\begin{eqnarray}
\delta \Omega[h_{i}] &=& \left(\gamma_{sl}-[\gamma_{lv}+\gamma_{sv}]\right)A_{sl}^{\rm bottom}\label{eq:omega_bottom} \\
\frac{\delta \Omega[h_{i}]}{V_{tot}}&=& \frac{1}{lh}\int_{0}^{l} {\rm d}x\;g(h_{i}(x))
\end{eqnarray}
with the interface potential $g(h_{i}) = - \gamma_{lv}(1+\cos
\theta_{Y})f(h_{i}/\delta)$. This introduces an additional length scale,
$\delta$, which characterizes the interaction range between the liquid-vapor
interface and the bottom of the cavity. In the absence of long-range forces,
this length scale is set by the width of the liquid-vapor interface, and we use
a value that is smaller than all other length scales of our macroscopic model.
In the numerical calculations we employ the following shape of the
interface potential
\begin{equation}
f(x) = \left \{ \begin{array}{ll} \left(1-x^{2}\right)^{2} & \mbox{for}\,0 \leq x\leq 1 \\ 0 & \mbox{otherwise}\end{array}\right.
\end{equation}
which smoothly interpolates between $0$ and $1$.

Thus, the excess grand potential is given by the following functional of
the interface position, $h_i(x)$
\begin{eqnarray}
\frac{\Delta \Omega[h_{i}]}{V_{tot}\Delta p_{\max}} &=& \Delta \tilde p \frac{1}{lh} \int_{0}^{l} {\rm d}x\;h_{i}(x)
- \frac{1}{2h\cos \theta_{Y}}  \int_{0}^{l} {\rm d}x\; \sqrt{1+\left(\frac{{\rm d}h_{i}}{{\rm d}x}\right)^{2}} \nonumber \\
&&- \frac{h_{i}(0)+h_{i}(l)+l}{2h}
+ \frac{1+\cos \theta_{Y}}{2h\cos\theta_{Y}} \int_{0}^{l} {\rm d}x\;f(h_{i}(x))
\end{eqnarray}
Rescaling the spatial coordinate by the width of the cavity, $\tilde x = x/l$,
and the position of the interface by the height of the cavity, $\tilde h(x) =
h_{i}(x)/h$, we arrive at the final expression
\begin{eqnarray}
\frac{\Delta \Omega[\tilde h]}{V_{tot}\Delta p_{\max}} &=& \Delta \tilde p \int_{0}^{1} {\rm d}\tilde x\;\tilde h(x)
- \frac{1}{2\cos \theta_{Y}}  \int_{0}^{1} {\rm d}\tilde x\; \sqrt{\alpha^{2}+\left(\frac{{\rm d}\tilde h}{{\rm d}\tilde x}\right)^{2}} \nonumber \\
&&- \frac{\tilde h(0)+\tilde h(1)+\alpha}{2}
+ \frac{\alpha(1+\cos \theta_{Y})}{2\cos\theta_{Y}} \int_{0}^{1} {\rm d}\tilde x\;f(\tilde h(\tilde x))
\end{eqnarray}
with $\alpha=l/h$, and $\delta$ is also measured in units of $h$. 

In the numerical calculations, the function $\tilde h(x)$ approximated by $N$
values, $\tilde h_{k}=\tilde h(\frac{k}{N-1})$ for $k=0, \cdots, N-1$. Thus,
the excess grand potential becomes a function of the $N$ rescaled interface
positions, $\tilde h_{k}$
\begin{eqnarray}
\frac{\Delta \Omega(\{\tilde h_{k}\})}{V_{tot}\Delta p_{\max}} &=& 
  \frac{\Delta \tilde p}{N-1} \left( \frac{\tilde h_{0}}2  + \sum_{k=1}^{N-2} \tilde h_{k} + \frac{\tilde h_{N-1}}2 \right) 
- \frac{1}{2\cos \theta_{Y}} \sum_{k=0}^{N-2} \sqrt{ \left( \frac{\alpha}{N-1}\right)^{2} + \left(\tilde h_{k}-\tilde h_{k-1} \right)^{2}}  \nonumber \\
&& - \frac{\tilde h_{0}+\tilde h_{N-1}+\alpha}{2}
+ \frac{\alpha(1+\cos \theta_{Y})}{2(N-1)\cos\theta_{Y}}  \left( \frac{f(\tilde h_{0})}2  + \sum_{k=1}^{N-2} f(\tilde h_{k}) + \frac{f(\tilde h_{N-1})}2 \right)
\end{eqnarray}

Using this explicit expression, we compute the chemical potential $\tilde
\mu_{k} \equiv \frac{1}{V_{tot}\Delta p_{\max}}\frac{\partial \Delta
\Omega}{\partial \tilde h_{k}}$
\begin{eqnarray}
\tilde \mu_{k} &=&  \frac{\Delta \tilde p}{N-1} 
- \frac{1}{2\cos \theta_{Y}} \left( \frac{\tilde h_{k}-\tilde h_{k-1}}{\sqrt{ \left( \frac{\alpha}{N-1}\right)^{2} + \left(\tilde h_{k}-\tilde h_{k-1} \right)^{2}} } + \frac{\tilde h_{k}-\tilde h_{k+1}}{\sqrt{ \left( \frac{\alpha}{N-1}\right)^{2} + \left(\tilde h_{k}-\tilde h_{k+1} \right)^{2}} }\right) \nonumber \\
&& + \frac{\alpha(1+\cos \theta_{Y})}{2(N-1)\cos\theta_{Y}} \frac{{\rm d}f}{{\rm d}\tilde h_{k}} \qquad \mbox{for}\, 1\leq k \leq N-2\\
\tilde \mu_{0} &=&  \frac{\Delta \tilde p}{2(N-1)} 
- \frac{1}{2\cos \theta_{Y}} \frac{\tilde h_{0}-\tilde h_{1}}{\sqrt{ \left( \frac{\alpha}{N-1}\right)^{2} + \left(\tilde h_{0}-\tilde h_{1} \right)^{2}} } - \frac{1}{2}+ \frac{\alpha(1+\cos \theta_{Y})}{4(N-1)\cos\theta_{Y}} \frac{{\rm d}f}{{\rm d}\tilde h_{0}}
\end{eqnarray}
and a similar expression holds for $\tilde \mu_{N-1}$.

We use these expressions to calculate the minimum free-energy path that is a
string of interface positions $\tilde h$ defined by the condition that the
chemical potential perpendicular to the path vanishes. Unlike
Eq.~(\ref{eq:MFEP}), we do not include the Jacobian, $M$, of the transformation
from the microscopic degrees of freedom, $\{{\bm r}\}$, of the particle model
to the collective variables, $\{\tilde h_{k}\}$ because we did not explicitly
specify the transformation between the microscopic and collective variables. We
note, however, that (i) the free-energy maximum of the string without $M$
corresponds to the saddle point of the free-energy landscape and (ii) that the
string corresponds to a continuous evolution of the liquid-vapor interface and,
in particular, mimics a locally conserved dynamics of the coarse-grained
density.

\section{Comparison between CREaM and the interface string}
\label{sec:CREaM}

\begin{figure}
\centering
	\includegraphics[width=0.6\textwidth]{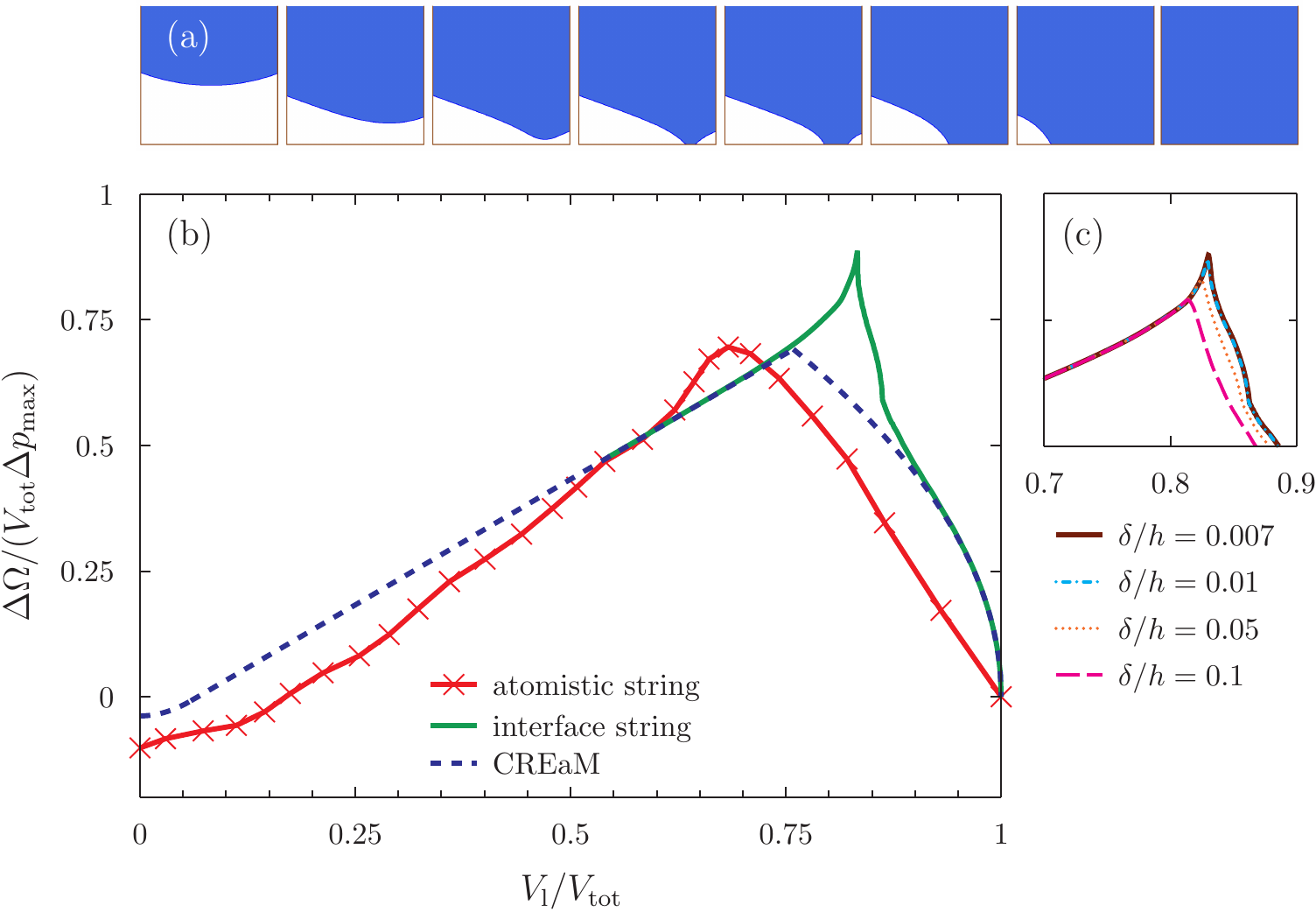}
	\caption{(a) Transition path computed via the interface string. (Multimedia view).
		(b) Rescaled excess free-energy along the interface string for $\Delta \tilde p=0$,
		$\alpha=1$, $\delta/h=0.0001$, $\theta_Y=110^\circ$, and $N=140$ (in green). The profile is
		compared to the atomistic and CREaM results at the same conditions. On the
		abscissa the parametric variable of the CREaM method $V_l/V_{tot}$ is reported;
		for the interface string $V_l/V_{tot}$ is computed from the path in the top
		panel, while for the atomistic string
		$V_l/V_{tot}=(Z-Z_{cassie})/(Z_{wenzel}-Z_{cassie})$, where $Z$ is the total
		number of particles inside the coarse-graining cells.  The saddle point
		corresponds to a configuration where the liquid-vapor interface touches the
		bottom of the capillary (between image 3 and 4 of the path above). (c) Behavior of the saddle point at different $\delta/h$.
		\label{fig:istring}}
\end{figure}

Here we discuss the relation between the wetting path identified by the
continuum rare events method (CREaM) previously introduced in Giacomello
\emph{et al.}\cite{Giacomello2012} and the interface ``string''.  We start by
introducing the committor function and isocommittor surfaces. The committor is
the probability that a trajectory passing by a state, an atomistic or a
macroscopic configuration, will reach next the product state,  say Wenzel,
rather than the reactant state. An isocommittor surface is the (hyper)surface
in the state space of constant committor function value. Thus, for example, the
isocommittor $0.5$ is the surface of states having $50$~\%  of probability to reach
first the products and $50$~\% to reach first the reactants. This surface is, by
definition, the transition state.

Consider an isocommittor surface, $S$ which  intersects the MFEP at a point
${\bm z}(\alpha)$. Let us denote by ${\bm n}_\alpha$ the normal to $S$ at
${\bm z}(\alpha)$, and by ${\bm \tau}({\bm z}_\alpha)$ the tangent to the MFEP
at the same point. In Ref.~\onlinecite{maragliano2006} it is shown that:
\begin{equation}
\label{eq:tangentVsNormal}
{\bm \tau}({\bm z}_\alpha) \parallel \hat M {\bm n}_\alpha.
\end{equation}
If we compare Eq.~\eqref{eq:tangentVsNormal} with Eq~\eqref{eq:MFEP}, we conclude
that the MFEP is the path connecting constrained minima of the free-energy on
the isocommittor surfaces $S_\alpha$ foliating the space.

Similar to the string method, CREaM aims at identifying a wetting path.
However, at variance with the string method, in CREaM this path is not
parametrized with its arc-length but with an observable of  relevance for the
problem at hand.  This very general framework can be applied to different
models, ranging from the micro- to the macroscale; for the macroscopic scale,
in the sharp-interface model that was originally used to describe the
Cassie-Wenzel transition, the wetting path is parametrized with the volume of
liquid in the groove, $V_l$. Like in the case of the
string method, in CREaM
the free-energy is minimized subject to the constraint that the variable
parametrizing the path is constant. In other words, the path obtained by CREaM
connects the constrained minima of the free-energy on the constant volume
surfaces $S^V_\alpha$.

Comparing this formulation of the CREaM path with the formulation of the string
path in terms of constrained minimization of the free-energy in the
isocommittor surface, we conclude that the two paths coincide if and only if
$S^V_\alpha \equiv S_\alpha$, at least locally to the path.

\begin{figure}
	\includegraphics[width=0.8\textwidth]{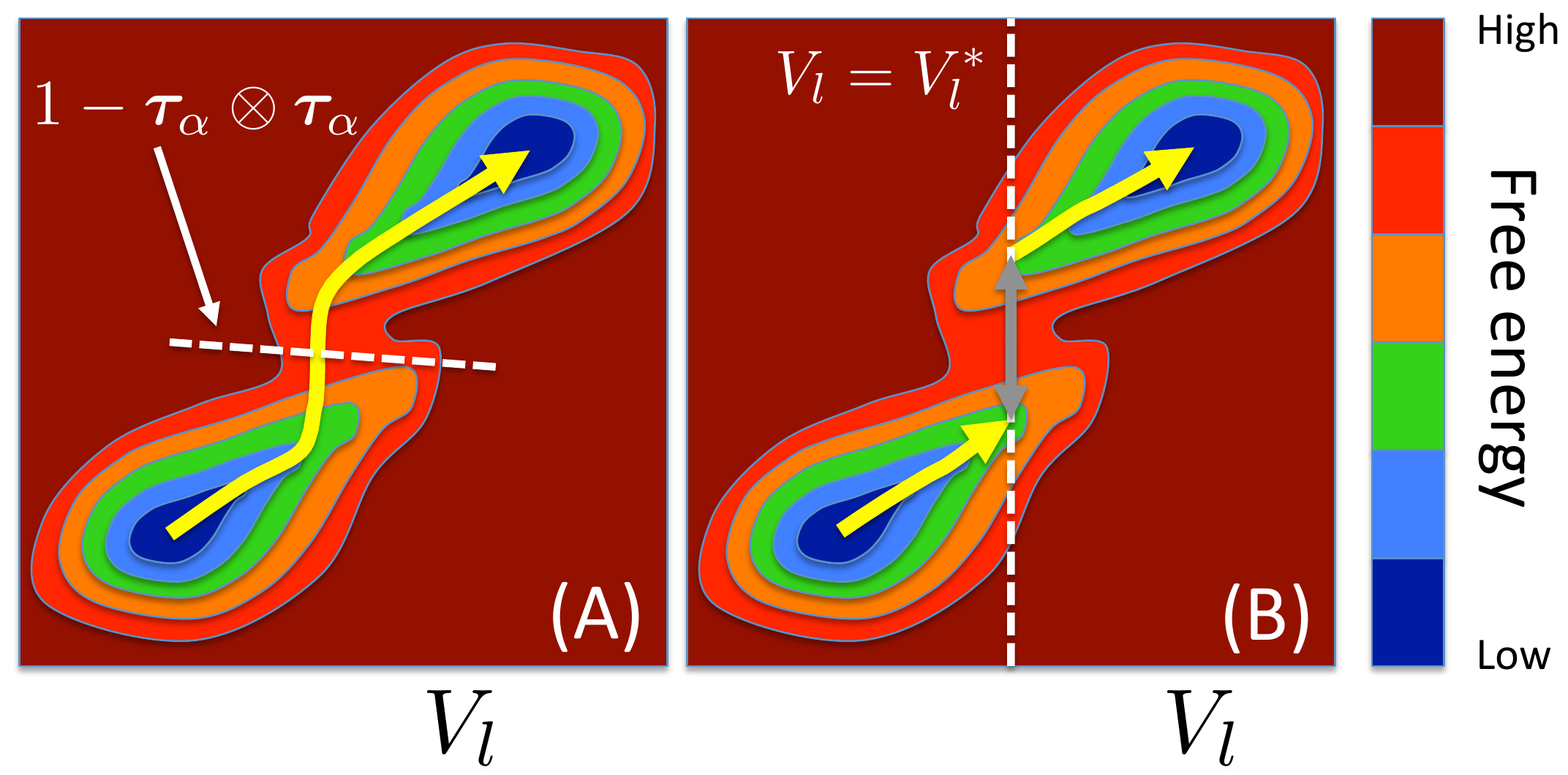}
	\caption{String (A) \emph{vs} CREaM (B) paths. In this sketch one of the axis represents the volume of liquid in the cavity $V_l$ while the other represents the complementary degrees of freedom. The dashed line in panel (A) is the hyperplane $\left ( \bm 1 - \bm \tau(\bm z(\alpha_l, t)) \otimes \bm \tau (\bm
z(\alpha_l, t)) \right ) \bm v$, where $\bm v$ is a generic vector. This is the plane on which $\hat M \nabla F$ is zero. The dashed line in panel (B), instead, represents the hypersurface $V = V_l$, i.e. one of the infinite hypersurfaces on which the minimum of the free energy is sought in the CREaM method.
	The string path
		follows the valley of the reactants and then moves smoothly to the valley
		of the products. When the reactants and products valleys become parallel, the
		CREaM path moves abruptly from one to the other, following the minimum of the
		free-energy corresponding to a given level of progress of the reaction. This is
		shown on the panel (B): the yellow line denotes the two branches of the CREaM
		path, and the gray double arrow highlights its discontinuity.
		\label{fig:CREaMvsStringPaths}}
\end{figure}

In Fig.~\ref{fig:istring}b we compare the  CREaM and interface string free
energy profiles along the respective paths. We remark that the profiles
coincide for most of the path, departing from each other only in a relatively
small region around the transition state. This is not surprising because, at
variance with the string, CREaM does not impose the continuity of the path.
Thus, in the case of the Cassie-Wentzel transition, which takes place through a
morphological transition (see Ref.~\onlinecite{Giacomello2012} and
Sec.~\ref{sec:results}), CREaM does not map the continuous path, along which the
symmetric meniscus configuration goes into the gas-bubble-in-a-corner one (see
Fig.~\ref{fig:CREaMvsStringPaths}). In Fig.~\ref{fig:istring}b we report also
the free-energy profile obtained from atomistic simulations. The agreement
seems to be better between atomistic and CREaM results than with interface string. The reason for this is discussed more in detail in the results
section; here we remark that this better agreement is due to ``errors
cancellation'', with the underestimation of the barrier in CREaM compensating
for an overestimation intrinsic to the sharp-interface models.

Summarizing, the paths obtained from CREaM and the string method are not
identical but give the same qualitative description of the process.  While the
string method gives a detailed and continuous description of the most likely
wetting path all along the process, CREaM represents the segment around the
transition state as a sharp morphological transition
(Fig.~\ref{fig:CREaMvsStringPaths}). Indeed, CREaM and the string can be used
as complementary tools. CREaM allows to efficiently compute all the possible
``reactive'' channels. The string method can then be used to further refine the
CREaM paths. When the system is relatively simple, like the case of wetting of
a square groove,~\cite{Giacomello2012}  it is possible to obtain the
analytical solution of the CREaM equations. Thanks to CREaM it was possible to
derive an extended version of the Laplace equation, which relates the
liquid/gas meniscus curvature to the surface tension. This relation, that was
introduced for the first time in Ref.~\onlinecite{Giacomello2012}, is valid along
most of the wetting path, apart in the region connecting the symmetric meniscus
and gas-bubble-in-a-corner morphologies. Finally, CREaM has a high parallel
efficiency, even higher than the string method as it does not require any exchange of data among images, and can be run on non connected heterogeneous computers.


\section{Results and Discussion}
\label{sec:results}
In this section, we present the ``physical'' results obtained \emph{via} the
atomistic string and the sharp-interface calculations (string and CREaM)
introduced in the previous sections.  We focus on the transition path for the
Cassie-Wenzel transition and on the related free-energy profiles. The length
scales covered range from few particle diameters $\sim11\sigma$ of the smallest
atomistic system simulated to macroscopic scales, which are described in terms
of sharp-interface models.

\subsection{The atomistic string}
\paragraph{The mechanism of the Cassie-Wenzel transition}

\begin{figure}
	\includegraphics[width=0.6\textwidth]{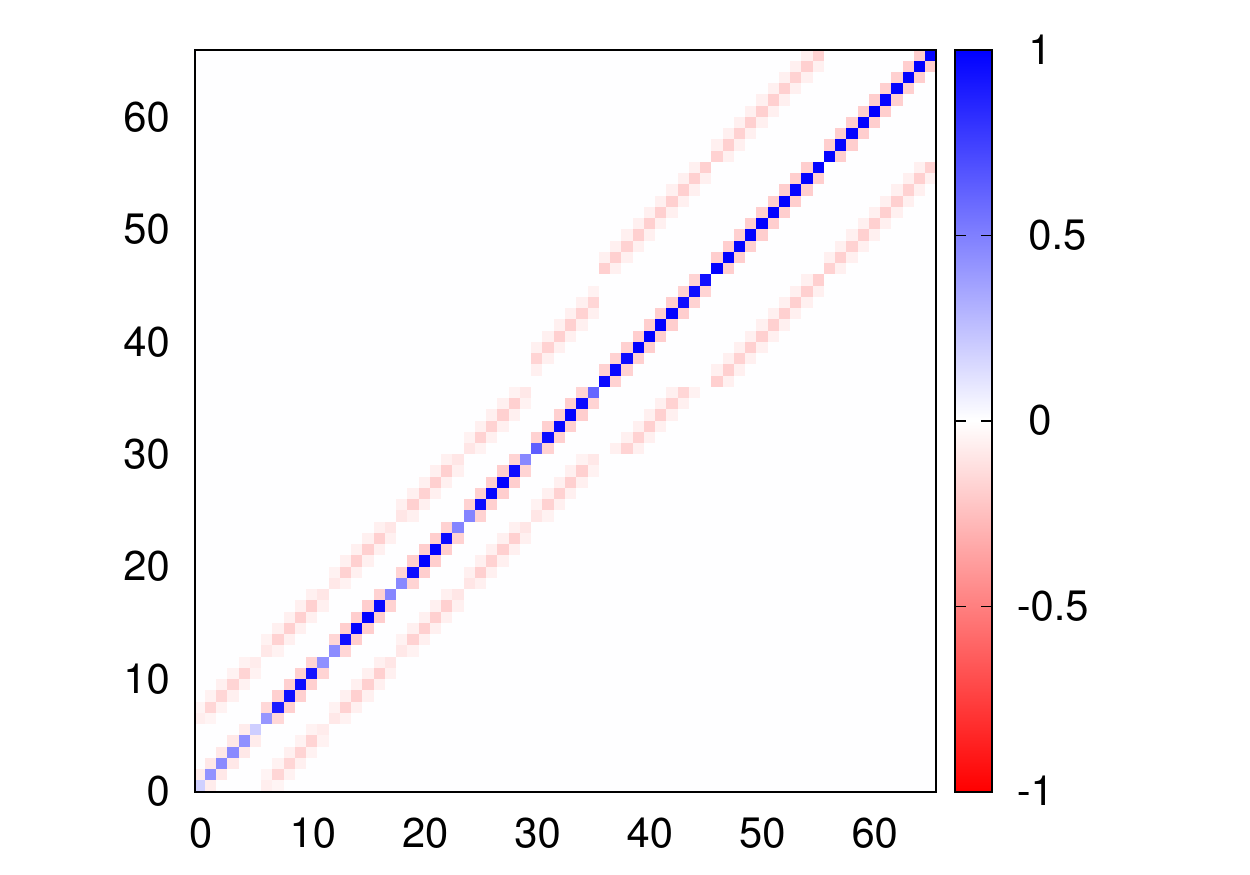}
	\caption{Metric matrix computed from Eq.~\eqref{eq:metricmatrix} at
		image $31$ (Wenzel state). The matrix elements are normalized
		with the maximum value. On the $x$ and $y$ axes is reported the cell
		number. Values lower than one are observed on the diagonal where the
		density is lower than the bulk liquid one (at the wall corners, see
		Fig.~\ref{fig:path}).
	\label{fig:MM}}
\end{figure}

We computed the transition path of the Cassie-Wenzel transition on two
geometries, a square and a rectangular groove (Fig.~\ref{fig:CV}).  A total of
$32$ images was used to discretize the string: the images and graphs that
follow are labeled with the image number. The pressure of the NPT simulations
was chosen to be close to the coexistence between the Cassie and the Wenzel
states.  The strings were initialized from configurations extracted from RMD
simulations with a single collective variable, number of particles in the
groove, analogous, apart for the ensemble (here NPT), to those presented  and
discussed in Ref.~\onlinecite{giacomello2012langmuir}.  We ensure that all initial
images in the string have the same symmetry, that is, all menisci lie in the
same corner, the left one. 

A general result of the atomistic string calculations is the form of the metric matrix along the string. A representative one is reported in Fig.~\ref{fig:MM}, showing that the most significant elements are those on the main diagonal. This result supports the assumption of a unitary metric matrix as is usually done in macroscopic, sharp-interface models. However, there are other very small but non-zero elements related to the surrounding coarse-graining cells as reflected by the multi-diagonal character of the metric matrix. These fine details could be encompassed in macroscopic models once the metric matrix is known from atomistic simulations. The detailed analysis of the effect of the metric matrix is deferred to a future study. 

\begin{figure}
	\includegraphics[width=\textwidth]{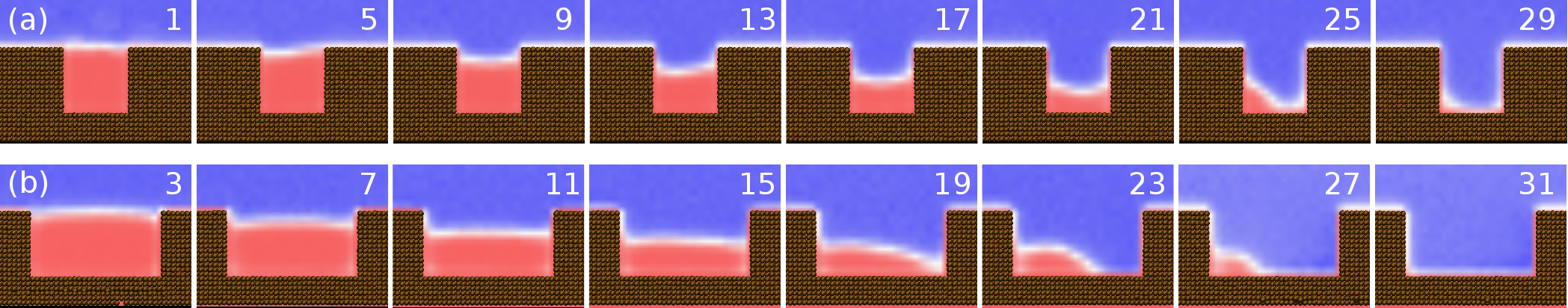}
	\caption{MFEP computed with the atomistic string method for grooves
	having square (a) and rectangular aspect ratios (b). The actual width of the
	rectangular groove is double the square one. Blue identifies high density,
	close to the liquid bulk one, while red low	density (vapor).  The image
	number is indicated in the corresponding density field. (Multimedia view). \label{fig:path}}
\end{figure}

The square groove measures around $11\sigma\times 11\sigma$. The
thermodynamic conditions were $T=0.8$ and $P=0.001$ in LJ units, where $P$ is the
global pressure observable computed for all atoms. Around $30$ steps of
evolution of the string (for each of which the mean forces were computed
\emph{via} RMD simulations, see Sec.~\ref{sec:algorithm}) were required to ensure convergence.

In Fig.~\ref{fig:path}a
we show the transition path for the square groove, \emph{i.e.}, the
sequence of average density fields forming the string at convergence.  The
meniscus is initially flat close to the Cassie state.  As the transition
proceeds, the meniscus descends into the groove with constant curvature
(images $7$-$21$) until close to the bottom a liquid finger is formed on the
right side of the groove (images $22$-$24$). Eventually the liquid wets one
corner of the groove forming a circular bubble that gradually shrinks (images
$25$-$29$) until it is completely absorbed and the Wenzel state is reached
(images $30$-$32$).

The initial and final parts of the path (the initial pinning, the symmetric
meniscus, and the final bubble in the corner) are in fair agreement with
previous restrained molecular dynamics simulations
\cite{giacomello2012langmuir} and the macroscopic CREaM results
\cite{Giacomello2012} (see Fig.~\ref{fig:intro}). However, close to the
transition state the liquid-vapor interface forms a finger thus departing from
the constant curvature menisci prescribed by CREaM. This discrepancy is
explained by the interface string path which, close to the transition state,
exhibits a point of the meniscus with high curvature (similar to the atomistic
finger) that eventually touches the bottom wall creating a small and a large
bubble (see Fig.~\ref{fig:istring}a). The fine
details of the process, however, are not easy to compare, because the diffuse
nature of the atomistic interface tends to smear out sharp points and small
vapor domains; to this must be added that computational constraints limit the
number of images and coarse-graining cells in the atomistic string.

The rectangular groove measures around $22\sigma\times 11\sigma$ and is
therefore twice as wide as the square one. The thermodynamic conditions
of the NPT simulations were $T=0.8$ and $P=0$ for this case. More than
$30$ steps of string evolution were required for convergence. 

The MFEP  for the rectangular groove is shown in 
Fig.~\ref{fig:path}b. It is seen that
before the Cassie minimum the meniscus curvature is allowed to vary while the
triple line is pinned at the top corners of the groove (images $1$-$5$), as is
expected from the macroscopic Gibbs' criterion~\cite{Oliver1977}.  This is an
evidence that pinning happens also at the nanoscale, even though in a particle
description of the system the continuum concept of ``geometrical singularity''
(corners at the top of the groove, in the present case) has no meaning.
However, the string resolution (number of images) does not allow us to quantify
the range of contact angles for which pinning occurs. The intrusion into the
groove happens when a sufficiently large meniscus curvature is reached, around
image $5$. The meniscus bends towards one corner at images $14$-$17$, earlier
in the progress of the transition than in the case of the square groove. This
observation can be made more quantitative by considering the length of the
string up to the transition state and normalizing it with the total length of
the string concerned with the activated event, that is, $\alpha_{TS}=
(i_{TS}-i_{\mathrm{cassie}}-1)/(i_{\mathrm{wenzel}}-i_{\mathrm{cassie}}-1)$,
where $i$ is the image number; for the rectangular groove $\alpha_{TS}=14/26$,
while for the square one $\alpha_{TS}=19/26$.  The contact line touches the
bottom wall (on the right hand side) and recedes ``rapidly'' towards the
opposite corner while the contact line at the vertical wall on the left does
not move. As a consequence, during the shrinking process the bubble at the
corner starts with a slightly flattened shape and then tends to a circular one.
This mechanism seems a generic one for the contact with the wall, since the
interface string path for the square groove also shows that the left contact
line behaves as if it were ``pinned'' at the vertical wall -- although there is
no defect -- while the right contact line slides on the bottom wall (see
Fig.~\ref{fig:istring}a).

\begin{figure}
	\includegraphics[width=0.6\textwidth]{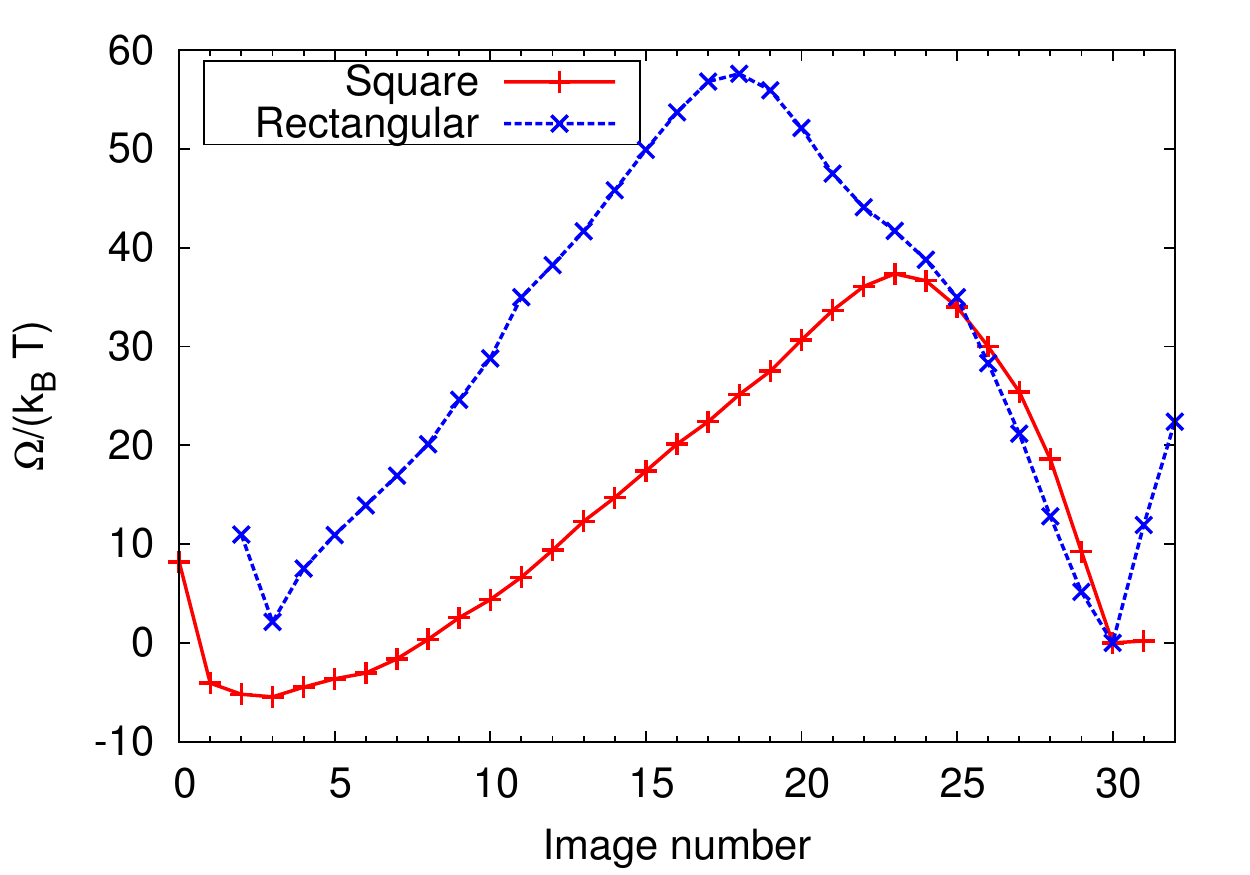}
	\caption{Free-energy profiles related to the MFEPs shown in
		Fig.~\ref{fig:path} in $k_BT$ units. The square groove is at
		$P=0.001$ and the rectangular groove at $P=0$. The image count for the
		rectangular groove is shifted by $+2$ in order to make the two Wenzel
		states coincide. The reference free-energy is taken to be that
		relative to the Wenzel state, $\Omega_W\equiv 0$.
	\label{fig:profiles}}
\end{figure}

The free-energy profiles related to the MFEPs just presented are shown
in Fig.~\ref{fig:profiles}. We remind that the profiles are at slightly
different pressures. In both cases, the free-energy barriers connected to the
Cassie-Wenzel transition are large as compared to the thermal energy
$k_B T$, a fact that corroborates the presence of strong metastabilities
on nano-rough hydrophobic surfaces. 
The free-energy barrier for the square groove is $34.3$~$k_B T$, while
for the rectangular one is $44.3$~$k_B T$, around $30\%$ larger,
suggesting that both the size and the aspect ratio of the groove have an
effect on the kinetics of the Cassie-Wenzel transition.

\begin{figure}
	\includegraphics[width=0.6\textwidth]{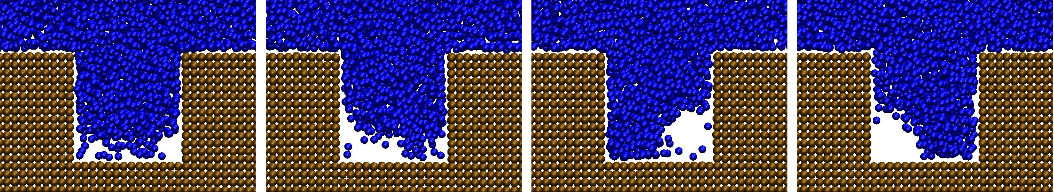}
	\caption{Atomistic configurations extracted from the transition state
	ensemble for the square nano-groove. (Multimedia view). \label{fig:TS_ensemble}}
\end{figure}

\paragraph{Validity of the concept of transition path}
\label{sec:TS_ensemble}
In Fig.~\ref{fig:TS_ensemble} we show atomistic configurations extracted
from the transition state ensemble of the square groove; these microstates were computed
\emph{via} RMD as detailed in Sec.~\ref{sec:TS}. It is apparent that the
microscopic configurations correspond to several macroscopic states:
bubble on the right corner, on the left one, and in the center. 
For the larger rectangular groove, instead, the transition state
ensemble is connected with a well defined configuration featuring a
bubble in the left corner, similar to that shown in Fig.~\ref{fig:path}b
(see also the related movie\footnote{See supplemental material at [URL will be inserted by AIP] for movies of the transition paths and the transition state ensemble.}). 

We ascribe this behavior to the flat free-energy landscape
along the hyperplane orthogonal to the transition state. When the barrier
separating the symmetry-related wetting paths (bubble-on-the-left and
bubble-on-the-right corners) is $\leq k_BT$ the system can easily jump
from one to the other.  In this case, the  wetting path identified by
the string (and CREaM) method has little statistical significance. In
other words, the transition
tubes~\cite{maragliano2006,vanden2006transition} around the two specular
paths, i.e., the region of state space in which most of the transition
trajectories pass through, overlap. Thus, we cannot describe the wetting
trajectories as ``fluctuations'' around the MFEP. In these conditions,
other methods, such as the finite temperature string~\cite{e2005}, would
be needed to capture information about the transition.

These results suggest that the macroscopic models of capillary systems
have a lower length scale below which they break down: the free-energy
landscape becomes too flat to identify a single \emph{macroscopic
state}. In the language of transition path theory, for sufficiently
small grooves the transition tube becomes large and the single
transition path found in the zero temperature limit has no statistical
significance. 

\subsection{The transition state}
\label{sec:TS_comparison}

\begin{figure}
	\includegraphics[width=0.6\textwidth]{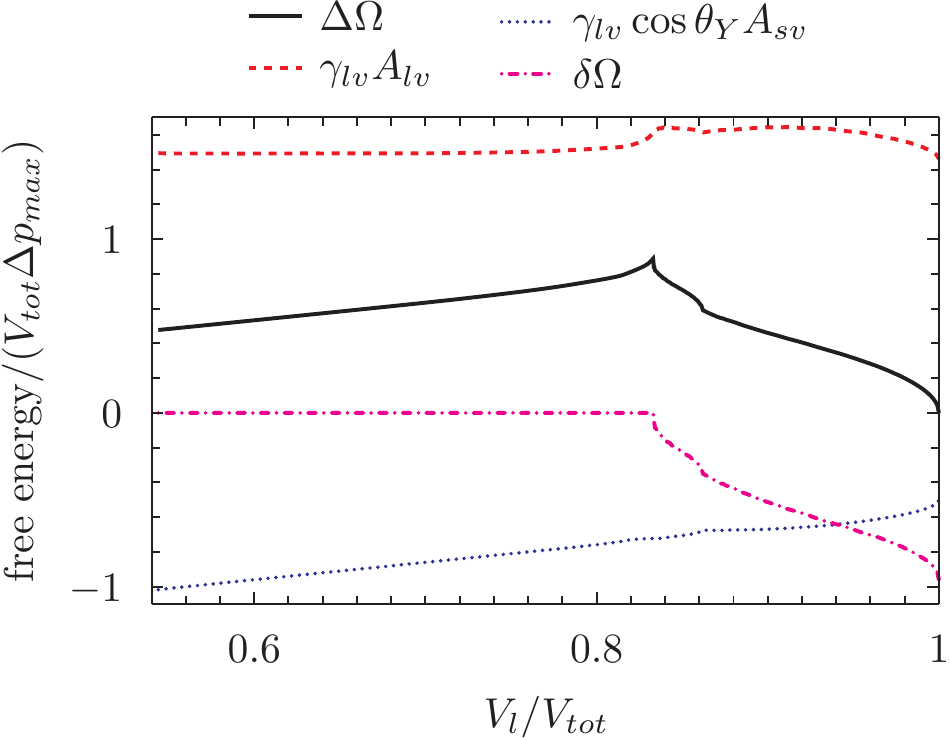}
	\caption{Contributions to the excess free-energy $\Delta \Omega$
	rescaled by $V_{tot} \Delta p_{max}$. The energetic costs of forming
	liquid-vapor or solid-vapor interfaces and of wetting the bottom wall are
	defined in Eqs.~\eqref{eq:omega_contrib} and \eqref{eq:omega_bottom},
	respectively. The results are computed via the interface string with $\Delta
	\tilde p=0$, $\alpha=1$, $\delta/h=0.0001$, $\theta_Y=110^\circ$, and $N=140$.
	\label{fig:energy_balance}}
\end{figure}

The interface string calculations were performed for the square groove in the
region of the transition state (the maximum of the profile), which is critical
both for evaluating the free-energy barriers and for testing the CREaM results.
As shown by the free-energy profiles in Fig.~\ref{fig:istring}b at low filling
levels ($V_l/V_{tot}<0.75$) and high ones ($V_l/V_{tot}>0.9$) the interface
string coincides with CREaM. Around the transition state, instead, the
interface string significantly departs from CREaM. The transition state itself
coincides with a cusp in the free-energy profile, while in CREaM it is a
non-differentiable point (see Fig.~\ref{fig:intro}). Ironically, the jump
discontinuity in the derivative at the transition state found with CREaM
induced us to further investigate the phenomenon, which is actually more severe
in the full-fledged string which shows an infinite discontinuity of the
derivative (in the sharp-interface limit $\delta\to 0$).  The interface string however guarantees the continuity of the
transition path, which is consistent with a continuous dynamics.  The cusp
arises because in touching the bottom wall the radius of curvature of the
meniscus has to change sign from  positive (symmetric meniscus) to negative
(asymmetric bubble). This is realized with a single point of zero radius of
curvature (and infinite curvature) developing on a side of the liquid-vapor
interface just before the contact with the bottom wall (see
Fig.~\ref{fig:istring}a). 
A different point of
view is that the area of the liquid-vapor interface significantly increases
during the formation of the liquid ``finger'' while $V_l$ remains almost
constant, thus giving rise to the sharp increase of the free-energy at the
transition state as clearly shown by the energy balance in
Fig.~\ref{fig:energy_balance}. The thermodynamic force $\mu$ corresponding to
the finger becomes infinite at the transition state. This divergence of $\mu$
is integrable, thus  giving rise to the cusp in the free-energy profile of
Fig.~\ref{fig:istring}b. 

In all cases analyzed -- atomistic, interface string, and CREaM -- the
transition state is connected with the contact of the  liquid domain with the
bottom wall of the groove (compare the paths in Figs.~\ref{fig:path},
\ref{fig:istring}a,  and \ref{fig:intro} with the free free-energy profiles in
Figs.~\ref{fig:istring}b): while the meniscus descends symmetrically in the
groove, the free-energy grows steadily because of the substitution of
vapor-solid interface with liquid-solid one which is unfavorable for
hydrophobic materials (see Fig.~\ref{fig:energy_balance}); when contact of the
meniscus with the bottom wall eventually occurs, new liquid-solid interface
replaces the liquid-vapor and solid-vapor interfaces at the bottom, resulting
in an overall reduction of the free-energy (the negative term $\delta \Omega$
in Fig.~\ref{fig:energy_balance}).

In the atomistic string, the transition state is found around one-half of the
string for the rectangular groove while it is towards the end of it for the
square one (see Fig.~\ref{fig:profiles}). The location of the transition thus
depends on the aspect ratio of the groove; this fact is easily explained by the
balance of the energy contributions above: for a taller groove the (growing)
branch of the free-energy connected with the meniscus sliding on the vertical
walls of the groove is longer, while the descending branch due to the shrinking
bubble is steeper.

The atomistic free-energy profiles reported in Fig.~\ref{fig:profiles} are
smooth at the transition state, differently from those obtained \emph{via} the
interface string and \emph{via} CREaM (see Fig.~\ref{fig:istring}b). In the atomistic case, indeed, thermal
fluctuations tend to smear out the extreme events seen in the sharp-interface
models. In particular, the formation of a point in the meniscus with very high
curvature is impossible in an atomistic picture. 
In order to account for this effect, we repeat the interface string calculations at increasing values of $\delta$, which correspond to wider liquid-vapor interfaces, see Fig.~\ref{fig:istring}c. The case $\delta/h=0.1$ roughly corresponds to the atomistic one, where the ratio of the interface thickness and the groove width is also $\sim 0.1$. Figure~\ref{fig:istring}c demonstrates that in the case of diffuse interfaces the transition state is smooth and the height of the free-energy barrier tends to decrease.

\section{Conclusions}
\label{sec:conclusions}
The atomistic string method in the density field collective variable was
applied for the first time to the Cassie-Wenzel transition to determine
rigorously the mechanism of the transition. The results of this work are both
methodological and physical. From the methodological point of view, we
demonstrated the relationship between approximate macroscopic
methods\cite{Giacomello2012} and the full-fledged interface string.  The former
methods are algorithmically simple and computationally convenient but fail
where more parallel valleys are present in the free-energy landscape.

The string simulations also offered physical insight into the mechanism of the
Cassie-Wenzel transition from the nanoscale to the macroscale. A morphological
transition was observed during the rare event with the meniscus changing from a
symmetric to an asymmetric configuration. The contact of the meniscus with the
bottom wall determines the position and shape of the transition state; at the
nanoscale the transition state is smooth, while in macroscopic models it shows
a cusp-like behavior.  The free-energy barriers are large compared to $k_B T$
even in nanoscale grooves; the depth of the groove and its aspect ratio are the
critical parameters to determine the kinetics of the Cassie-Wenzel transition.
It was also shown that for very small grooves (width $\sim 11\sigma$) the
concept of transition path breaks down and it is not possible to identify a
unique sequence of macroscopic configurations that describe the Cassie-Wenzel
transition.

\begin{acknowledgements}

The research leading to these results has received funding from
the European Research Council under the European Union's
Seventh Framework Programme (FP7/2007-2013)/ERC Grant agreement n. [339446].  
S.M. acknowledges financial support from the MIUR-FIRB Grant No. RBFR10ZUUK.
M.M. thanks the SFB 803 (B03) for financial support.
We acknowledge PRACE for awarding us access to resource
FERMI based in Italy at Casalecchio di Reno.

\end{acknowledgements}


%

\end{document}